# Microscopic Description of Thermodynamics of Lipid Membrane at Liquid-Gel Phase Transition

Boris Kheyfets[1], Timur Galimzyanov[1,2] and Sergei Mukhin[1]

[1]: National University of Science and Technology "MISIS"

[2]: A.N. Frumkin Institute of Physical Chemistry and Electrochemistry, Russian Academy of Sciences

**Abstract**: A microscopic model of the lipid membrane is constructed that provides analytically tractable description of the physical mechanism of the first order liquid-gel phase transition. We demonstrate that liquid-gel phase transition is cooperative effect of the three major interactions: inter-lipid van der Waals attraction, steric repulsion and hydrophobic tension. The model explicitly shows that temperature-dependent inter-lipid steric repulsion switches the system from liquid to gel phase when the temperature decreases. The switching manifests itself in the increase of lateral compressibility of the lipids as the temperature decreases, making phase with smaller area more preferable below the transition temperature. The model gives qualitatively correct picture of abrupt change at transition temperature of the area per lipid, membrane thickness and volume per hydrocarbon group in the lipid chains. The calculated dependence of phase transition temperature on lipid chain length is in quantitative agreement with experimental data. Steric repulsion between the lipid molecules is shown to be the only driver of the phase transition, as van der Waals attraction and hydrophobic tension are weakly temperature dependent.

## 1. Introduction

Over the years main phase transition in lipid membrane has been treated with various theoretical approaches, such as phenomenological [1], semi-emperical [2], and a number of microscopic approaches [3–5]. Some of the microscopic approaches resort to numerical methods [4,5], though so far obtained results differ substantially from experiment [3]. Hence, consistent and comprehensive analytical description of the main phase transition is still missing. This paper is aimed to close this gap by considering consistently lipid molecules steric repulsion, van der Waals (VdW) attraction, and hydrophobic tension at the membrane water interface. The novelty of our approach is application of path-integral technique for description of steric repulsion between hydrocarbon chains.

Liquid to gel phase transition, referred to as "main phase transition" in literature, is first-order phase transition [6,7] characterized by higher ordering of the lipid tails [8], a reduction of area per lipid and increase in membrane thickness [8,9], a growth by an order of magnitude of membrane rigidity [10], a drop in self-diffusion coefficient [11], and change of volume per lipid [12]. Recently, main phase transition has been a topic of several molecular dynamics studies [8,13]. Though, for most lipids a corresponding sin-gle-component lipid membrane undergoes this transition at a temperature significantly lower than the room temperature, an important exceptions exist for lipids with fully saturated hydrophobic chains e.g. DMPC ($T_m$ = 24 °C) and DPPC ($T_m$ = 41 °C). Moreover, a number of recent studies suggest that cholesterol might *induce* gel phase in lipid membrane [14,15].

Present study belongs to a new stream of works caused by the recent increase of interest to the membrane thermodynamics in a view of its role in the functioning of proteins [16–19].

In this paper we consider microscopic model and present a theory that makes transparent the physical mechanism of the liquid-gel phase transition. Our calculation results, obtained in a single approach, describe corresponding changes in the various thermodynamic characteristics, as well as dependence of main transition temperature on chain length. Though molecular dynamics studies [8,13] achieved quantitative agreement with experiment in each particular case, they serve rather as numerical experimental tool complementing generalized physical concept of the underlying mechanisms clarified by the theory.

One of the key features that distinguishes lipid membranes from conventional solids is the ability of lipid molecules to self-assemble into various structures in water. This ability is due to amphiphilic nature of lipid molecules and water being a polar solvent. Water results in a hydrophobic tension at the interface of lipid bilayer with water, which is balanced by steric repulsion of lipid chains. In the simplest model the energy of the steric repulsion is inversely proportional to area per lipid, $A$, while the energy of the hydrophobic tension is proportional to $A$ [20,21]: $E(A) = \gamma A + \alpha/A$. These two contributions *alone* give the self-assembling of the lipids, manifested by finite equilibrium area per lipid at $A_0 = \sqrt{\alpha/\gamma}$. However, the liquid-gel phase transition cannot be described with this simple model because in this model the energy as a function of area per lipid, $E(A)$, possesses only one single minimum (at $A_0$). On the other hand, a first-order phase transition requires at least two energy minima of equal energies at different areas, between which the transition occurs. Addition of van der Waals attraction between the lipid chains [20] allows an adequate description of the thermodynamic properties of lipid membranes above and below transition temperature, $T_m$. In this paper we add van der Waals attraction into our microscopic model of lipids [22,23]. Thus obtained model describes the liquid-gel phase transition. The model allows for steric repulsion between lipid chains via self-consistent recasting the entropy of their different conformations, each chain being modeled with semi-flexible string. The hydrophobic tension is left in its simplest form [20]. For the reference we use experimental data for 1,2-dipalmitoyl-sn-glycero-3-phosphocholine (DPPC) lipids.

## 2. Calculation

### 2.1. Flexible strings model

We briefly recap the flexible strings model [22,23]. Layers of a bilayer membrane are assumed to slide freely with respect to each other. Lipid of a single monolayer is modeled as an effective flexible string with a given incompressible area, $A_n$, and finite bending rigidity, $K_f$, subjected to the confining parabolic potential. The latter allows for a repulsive entropic force induced between the neighboring lipid molecules in the same monolayer, due to excluded volume effect. Interaction between heads is effectively included into surface tension in the hydrophobic region. Bending energy of the effective string is an ad-hoc way of taking into account trans-grauch chain conformations. Energy functional of the string consists of kinetic energy and bending energy of a given dynamical string conformation, as well as potential energy in the confining potential induced by collisions with the neighboring strings:

$$E_t = \int_0^L \left[ \frac{\rho \dot{\mathbf{R}}^2}{2} + \frac{K_f}{2} \left( \frac{\partial^2 \mathbf{R}}{\partial z^2} \right)^2 + \frac{B \mathbf{R}^2}{2} \right] dz , \qquad (1)$$

here $\rho$ is a string linear density, $\mathbf{R}(z) = \{R_x(z), R_y(z)\}$ is a vector in the plane of the membrane giving deviation of a string from the straight line as a function of coordinate $z$ along the axis normal to the membrane plane, and $B$ is a parameter of the confining potential determined self-consistently. Self-consistent parabolic potential has been used previously to model a polymer chain in confined geometry [24]. That approach is conceptually close to the statistical kink-model [25,26], which was used to find probability distribution function of chain conformations and macroscopic membrane characteristics, minimizing free energy of the membrane. In contrast to that model we use continuous description of the lipid chain bending fluctuation and include an option of the direct self-assembly of lipids in membrane.

We use the same boundary condition and approach, as in original paper [22]. Boundary conditions for a model flexible string take into account the following physical assumptions (same is assumed also for component $R_y(z)$): $R'_x(0) = 0$ – a chain terminates perpendicularly to the membrane surface, $R'''_x(0) = 0$ – net force acting on a head is zero, $R''_x(L) = 0$ – net torque acting at a chain's free end is zero, $R'''_x(L) = 0$ – net force acting at a chain free end is zero. The first boundary condition reflects the orientational asymmetry of the monolayer due to water-lipid interface which is clearly seen from data on the molecules orientational order parameter [27,28]: lipid tails are more ordered in the vicinity of headgroups constrained by the hydrophobic tension. Yet, the chains are not permanently perpendicular to the membrane surface, and boundary conditions are approximate and necessary to keep the model analytically solvable.

Assuming membrane to be locally isotropic in the lateral plane, partition function can be split in the product of the two equal components, $Z = Z_x Z_y = Z_x^2$, and thus free energy of the lateral oscillations of the chain equals to:

$$F_t = -2k_B T \log Z_x \qquad (2)$$

Partition function $Z_x$ could be written as a path integral over all chain conformations:

$$Z_x = \iint e^{-\frac{E\{R_x(z), \dot{R}_x(z)\}}{k_B T}} DR_x D\dot{R}_x, \qquad (3)$$

which can be easily integrated due to the quadratic form of the energy functional (1) [22,23].

To derive the self-consistency equation for so far unknown parameter *B*, we differentiate both sides of the Eq. (2) with respect to *B* and readily obtain the self-consistency equation for this parameter:

$$\frac{\partial F_t}{\partial B} = L \langle R_x^2 \rangle \qquad (4)$$

where brackets denote thermodynamic (Boltzmann) average over chain conformations. R.h.s. of Eq. (4) is directly expressed via the thermodynamic average area per lipid *A* in the membrane plane and effective incompressible area of lipid chain $A_n$:

$$\pi \langle R_x^2 + R_y^2 \rangle = 2\pi \langle R_x^2 \rangle = \left( \sqrt{A} - \sqrt{A_n} \right)^2 \qquad (5)$$

where an isotropy of the membrane in $x, y$ plane is assumed for simplicity. Solving this equation one finds *B(A)* (for more details see [29] and *Supplementary Materials A*).

Free energy of the string is equal to the sum of lipid tail free energy Eq. (3) and hydrophobic tension energy $\gamma A$:

$$F_T = F_t + \gamma A \qquad (6)$$

Equilibrium area per lipid can be found by minimizing expression (6) over the area per lipid *A* in the membrane plane, which leads to the following balance equation for a membrane with zero external lateral pressure applied at the membrane's perimeter (self-assembly condition):

$$\frac{\partial F_T}{\partial A} = \frac{\partial F_t}{\partial A} + \gamma = 0 \quad \Rightarrow \quad \gamma = -\frac{\partial F_t}{\partial A} \equiv P_t \qquad (7)$$

Condition (7) simply means that in the equilibrium entropic repulsion of the chains, $P_t$, should be balanced by hydrophobic tension, i.e. the attraction of the heads. One obtains area per lipid in a membrane by solving self-assembly condition Eq. (7) with respect to *A*, with free energy $F_t$, Eq. (2).

In order to describe the temperature driven liquid-gel phase transition we add the van der Waals attraction between lipid hydrocarbon chains [20] to the membrane free energy in Eq. (6), and account for the dependence of the membrane thickness on the temperature. To derive dependence of the monolayer thickness, *L*, on the temperature we relate it to the contour length of the lipid chain $L_R$ proportional to the number of $CH_2$ groups in the lipid tail:

$$L_R = \int_0^L \sqrt{1 + \left\langle \left( \frac{\partial \vec{R}}{\partial z} \right)^2 \right\rangle} \, dz \qquad (8)$$

(for a derivation see *Supplementary Material B*).

Since a single lipid molecule with two hydrocarbon tails is modeled with a single semi-flexible string, the van der Waals attraction in our model is an effective interaction between different lipid molecules. We use a standard expression for this interaction, but with effective strength constant. Analytical expression for van der Waals interaction between two hydrocarbon chains has been calculated in [30] on the basis on quantum mechanics:

$$W = U \frac{3\pi}{8L} \frac{N^2}{D^5} \quad (9)$$

where $L$ is a length of the chain, $N$ is a number of $CH_2$ groups in a single chain, $D$ is a distance between the chains (see Fig. 1 in [30]), and $U$ is interaction constant. Associating $D$ with an average square root of the area per lipid molecule in a lateral plane, one arrives at the following equation for the lipid membrane free energy:

$$F_T = F_t(A) + \gamma A - \frac{\tilde{U} N^2}{L A^{5/2}} \quad (10)$$

where renormalized coefficient is:

$$\tilde{U} = U \frac{9\pi^{7/2}}{2^8} \quad (11)$$

an additional factor of 3 accounts for the fact that every lipid is surrounded by other lipids, assuming a tightly packed lattice, which has a coordination number 6.

## 2.2. The main phase transition

As mentioned in the *Introduction*, the simplest model for steric repulsion is $F_t \sim 1/A$. Taking into account the finite width of the hydrocarbon chain, we can rewrite it as $F_t \sim 1/(A - A_n)$; so that steric repulsions outgrows van der Waals attraction at smaller area per lipid. At lower temperatures, area per lipid is small, hence area per lipid is defined by the interplay between steric repulsion and van der Waals attraction. At higher temperatures, area per lipid is larger; with van der Waals attraction's strong dependence on area per lipid, $F_{VdW} \sim 1/A^{5/2}$, steric repulsion and hydrophobic tension are effectively the two main contributions to the area per lipid.

We normalize the free energy of lipid tails by the free energy of the rigid rods ($K_f \to \infty$) [31]. In that limit the expression for the free energy could be found explicitly by neglecting the string bending term in the energy functional Eq. (1). This leads to the expression for regularized free energy of the semi-flexible string:

$$\frac{\tilde{F}_t}{k_B T} = 2 \log \frac{\hbar \sqrt{\frac{K_f}{L^4 \rho}}}{k_B T} + \frac{3}{4} \log \left( L^4 \frac{B}{K_f} \right) + \sqrt{2} L \sqrt[4]{\frac{B}{K_f}} \quad (12)$$

(for a derivation see *Supplementary Materials C and D*).

Solving self-consistency Eq. (4) we than substitute dependence $B(A)$ into the expression for the total free energy $F_T$ (Eq. (10)), normalized by the free energy of the rigid rods (see Eq. (12)). Thus we find the dependence of the free energy on the area per lipid and temperature (see plotted in Fig. 1). The DPPC lipid bilayer is used as a reference, as one of the most characterized lipid. The choice of the microscopic model parameters: $L_R = 16$ Å, $A_n = 20$ Å², $K_f = 5.797 10^{-21}$ erg cm, and $U = 745$ kcal Å⁶/mol, are made to match respective DPPC micro- and macroscopic parameters, such as: main transition temperature and area per lipid in the vicinity of the transition. See *Discussion* for details on lipid parameters used in this paper and stability analysis. Our results are plotted in Fig. 1.

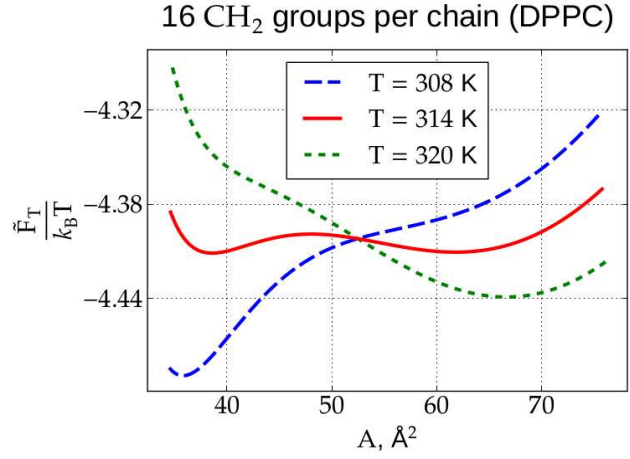

*Fig. 1: Free energy of the DPPC bilayer as a function of the area per lipid at three different temperatures: at $T_m$, above and below $T_m$.*

At low and high temperatures the system has only one minimum, corresponding to the area per lipid in the gel and the liquid phase respectively. Near the critical temperature the two minima corresponding to first order phase transition occur. For DPPC at $T = 320$ K and at $T = 308$ K we obtained area per lipid equal to 66 Å², and 36 Å² correspondingly. These results qualitatively match experimental ones (see Table 6 in [9]).

We calculate dependences on the area per lipid of all three terms contributing to the free energy, Eq. (10), at different temperatures (see Fig. 2).

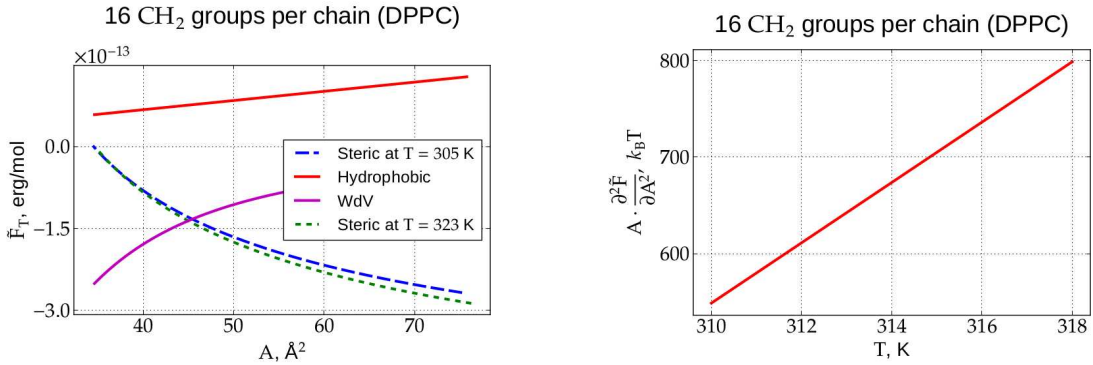

*Fig. 2: Left: Hydrophobic and VdW interactions do not depend on temperature (see Eq. (10)). Hence, steric repulsion is the one responsible for the main phase transition in lipid membranes. Right: Lipid membrane compressibility coefficient is increasing with temperature. This effect leads to the main phase transition in lipid membranes (see Fig. 2 (left) and Fig. 1).*

One can see that both hydrophobic and VdW interactions are responsible for membrane's compression, while steric interaction favors membrane's expansion. Among considered lipid interactions, only steric repulsion depends on temperature (as is seen from Eq. (10)). This means that steric interaction is responsible for the temperature-driven liquid-gel phase transition. An increase of temperature leads to steeper dependence of steric repulsion on the area per lipid. This is manifested by the temperature de-pendence of membrane compressibility modulus, $K_a = A \cdot \frac{\partial^2 \tilde{F}}{\partial A^2}$, (see Fig. 2, right). The modulus increases with temperature due to more frequent chain collisions, inducing stronger steric repulsion. Thus, at lower temperatures fluctuations are suppressed, hence, steric repulsion is reduced, which in turn favors the gel phase with lower area per lipid.

## 3. Thermodynamic properties

Good quantitative agreement with experimental data is obtained for dependence of phase transition temperature $T_m$ on the chain length (number of $CH_2$ groups) — see Fig. 3. Deviation of calculation results from the experimental data at high temperatures could be caused by the change of hydrophobic tension at the lipid-water interface due to close proximity of the temperature $T_m \sim 80$ °C to the water boiling temperature.

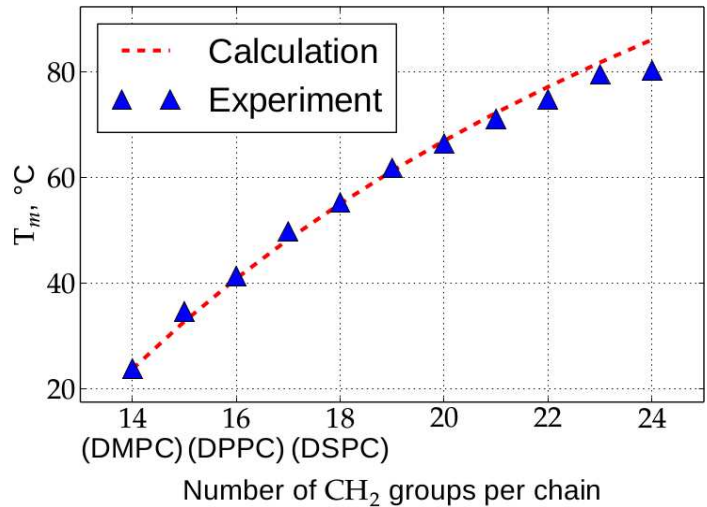

*Fig. 3: Comparison of calculated dependence of gel-liquid phase transition temperature $T_m$ on the chain length with experimental data [36–38]. Changing just the bare contour length, $L_R$, of the lipid in proportion to the number of $CH_2$ groups results in a good agreement with experiments. DMPC stands for 1,2-dimyristoyl-sn-glycero-3-phosphocholine; DSPC stands for 1,2-distearoyl-sn-glycero-3-phosphocholine.*

We have extracted from our theory the temperature dependencies of the membrane thickness $L$ from Eq. (8) and volume per $CH_2$ group $v_{CH_2} = LA/N_{CH_2}$ — see Fig. 4. These results are derived using calculated jump of the area per lipid, $A$, at the main transition, see Fig. 1. Our results semi-quantitavely match experimental data [9,12,32]. We calculated temperature dependence of the area per lipid for lipids with various chain lengths (see Fig. 4). The magnitude of the area jump at $T_m$ transition temperature increases with the lipid chain length in accord with experiment [33].

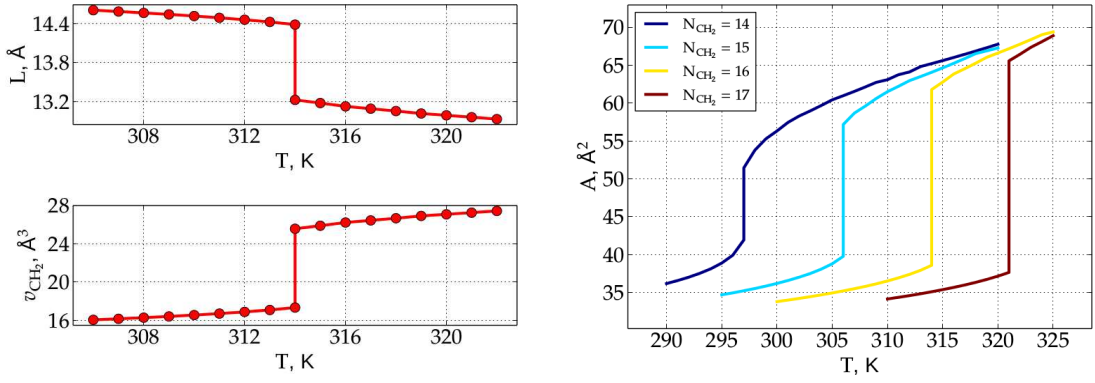

*Fig. 4: Upper left: Calculated dependence of thickness L of hydrocarbon part of DPPC monolayer on temperature. Lower left: Calculated dependence of volume per $CH_2$ group on temperature. Right: Calculated temperature dependences of area per lipid for lipids with different chain lengths.*

Our model is formulated in the coordinate system, in which $Oz$ axis is parallel to an average direction of a lipid chain. That means that at nonzero tilt angle, the calculated area $A$ is obtained in the plane perpendicular to the chain axes, rather than in the membrane plane. Therefore, the calculated area per chain in gel phase is smaller then the experimental ones due to tilt effect.

## 4. Discussion

In this paper we used microscopic model of lipid membrane, to reproduce liquid-gel phase transition with temperature. This is achieved by combining quantum mechanical calculation of Van der Waals attraction between hydrophobic chains [30], analytical calculation of steric repulsion [22], and linear model of hydrophobic tension [34].

This phase transition is usually described as straightening of the hydrophobic chains which, presumably, causes the increase of VdW interaction between neighboring chains. We point out that this picture is not directly self-consistent: increased chain length *decreases* the energy of the VdW interaction (see Eq. (9)), while corresponding change of the area per lipid depends, in principle, on the change of the volume per lipid molecule. We have demonstrated that inter-molecular entropic repulsion is the only term which directly depends on temperature (see Fig. 2). Its decrease with lowering the temperature shifts the free energy minima to smaller areas per lipid (see Fig. 1). Hence, liquid-gel phase transition should be treated as cooperative effect of at least three interactions: VdW, steric repulsion and hydrophobic tension.

The cooperative nature of liquid-gel phase transition is captured self-consistently in our calculation: as the temperature lowers, free-energy minimum shifts to the smaller area per chain, which in turn, results in the increase of steric repulsion parameter $B$ entering the confining potential, (see Eq. (1)).

We do not consider kinetics of phase transition in this manuscript and work in the thermodynamical limit, considering membrane in its equilibrium state. Thus our model is not applicable in the very close vicinity of the critical temperature. This region should be described within another approach specifically designed to describe critical phenomena from the ground up.

Our calculations are in a qualitative agreement with experimental data for area per lipid, and main transition temperature $T_m$ as functions of lipid chain length. Parameters of the flexible strings model are the following: incompressible area, bending rigidity of the string and hydrophobic tension of the membrane, and contour length of the chain. Despite the large number of fitting parameters they all are strongly confined to the physical parameters of the lipid molecules, so the model results are not easily tunable. Incompressible area, $A_n$, characterizes the entropic repulsion between neighboring chains. In the manuscript its value is taken to be 20 Å$^2$. Incidentally, this value is slightly above the crystalline area per lipid (18 Å$^2$). Bending rigidity of the chain has been estimated on the basis of elasticity theory — see note [7] in [23]. There we obtained an estimate: $K_f = 0.33$ $k_B$ T $L_0$, where $L_0$ is the characteristic thickness of the hydrofobic part of lipid monolayer. In this manuscript we had to choose a more rigid strings with $K_f = 0.836$ $k_B$ $T_0$ $L_0$ to fit the experimental results, which correspond with rough estimate ($T_0$ is room temperature, $T_0$=300 K; $L_0$ = 13 Å). Hydrophobic tension $\gamma$ is commonly estimated to be in the range of 20-50 erg/cm$^2$ [2,20,35]. In the manuscript we used a slightly lower value for $\gamma$ — 17 erg/cm$^2$. The total length of the chain is taken to be 16 Å which we find reasonable with given that this value doesn't include the length of the hydrophilic heads and the overall thickness of bilayer is about 2 nm.

Van der Waals interaction coefficient is estimated to be 1340 kcal Å$^6$/mol [30]. We obtained a two times smaller value of 745 kcal Å$^6$/mol, that gives theoretical transition temperature matching the DPPC data. The discrepancy might be caused by the difference between an idealized picture considered in [30] for calculation of the van der Waals interaction and complexity of lipid molecules in the actual bilayer.

Analysis of model prediction for the transition temperature shows that changing values of model parameters: $A_n$, $K_f$, $U$ (VdW parameter) by one percent result in $T_m$ changes of at most one percent (see *Supplementary Material E*).

## 5. Conclusion

We had developed analytical theory of the first order liquid-gel phase transition of the lipid bilayer using microscopic model of semi-flexible strings. We had demonstrated analytically that steric repulsion, hydrophobic tension and van der Waals interaction are essential for this phase transition, while steric repulsion is the driver of the transition, as it is the only factor of the these three which depends directly on the temperature. Membrane compressibility modulus increases with temperature inducing stronger steric repulsion and transition from gel to liquid phase with higher value of area per lipid and disorder. We applied our approach to calculate change in the multiple thermodynamical properties as a result of phase transition such as: area per lipid, volume per hydrocarbon group, and membrane thickness.

The qualitative picture captured by our theory is supported by quantitative correspondence with a number of experiments. The phase transition temperature dependence on the contour length of the lipid chain (number of the hydrocarbon groups) found theo-retically is in good quantitative agreement with experimental data for DSPC, DPPC and DMPC lipids (see Fig. 3). Many other properties might be considered within the same microscopic model: chains orientational order parameter, membrane bending modulus, gauss modulus, lateral pressure profile, etc.

## Acknowledgements

The authors are grateful to Peter Kuzmin for reading and discussing early version of the manuscript. The authors gratefully acknowledge the financial support of this work by RSCF grant (project no. 17-79-20440).

# Microscopic Description of Thermodynamics of Lipid Membrane at Liquid-Gel Phase Transition

Boris Kheyfets[1], Timur Galimzyanov[1,2] and Sergei Mukhin[1]

[1]: National University of Science and Technology "MISIS"

[2]: A.N. Frumkin Institute of Physical Chemistry and Electrochemistry, Russian Academy of Sciences

## A. Flexible strings model overview

Flexible strings model is treated within a mean-field theory, that considers lipid in a self-consistent field of other lipids in the same monolayer. Layers of a bilayer membrane are assumed to slide freely with respect to each other. Lipid of a single monolayer is modeled as an effective flexible string with a given incompressible area, $A_n$, and finite bending rigidity, $K_f$ (see Fig. 1), subjected to the confining parabolic potential. The latter allows for a repulsive entropic force induced between the neighboring lipid molecules in the same monolayer, due to excluded volume effect (see Fig. 2). Interaction between heads is effectively included into surface tension in the hydrophobic region. Energy functional of the string consists of kinetic energy and bending energy of a given dynamical string conformation, as well as potential energy in the confining potential induced by collisions with the neighboring strings:

$$E_t = \int_0^L \left[ \frac{\rho \dot{\mathbf{R}}^2}{2} + \frac{K_f}{2} \left( \frac{\partial^2 \mathbf{R}}{\partial z^2} \right)^2 + \frac{B \mathbf{R}^2}{2} \right] dz , \qquad (A.1)$$

here $\rho$ is a string linear density, $\mathbf{R}(z) = \{R_x(z), R_y(z)\}$ is a vector in the plane of the membrane giving deviation of a string from the straight line as a function of coordinate $z$ along the axis normal to the membrane plane (see Fig. 1), and $B$ is a parameter of the confining potential determined self-consistently. Self-consistent parabolic potential has been used previously to model a polymer chain in confined geometry [1]. That approach is conceptually close to the statistical kink-model [2,3], which was used to find probability distribution function of chain conformations and macroscopic membrane characteristics, minimizing free energy of the membrane. In contrast to that model we use continuous description of the lipid chain bending fluctuation and include an option of the direct self-assembly of lipids in membrane.

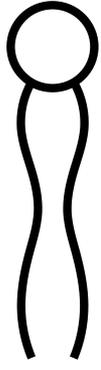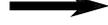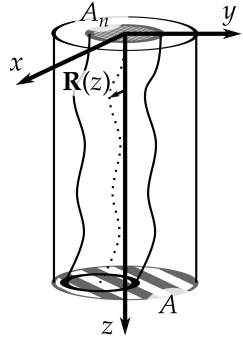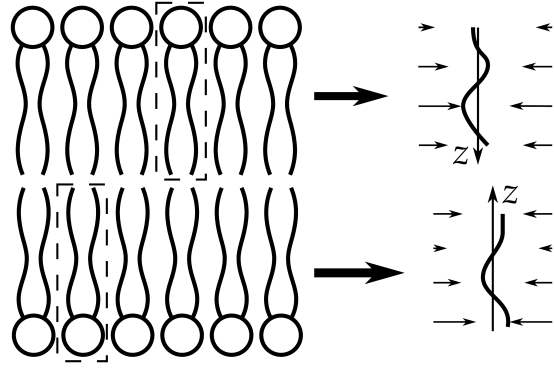

*Fig. 1: Hydrocarbon tails of lipid are modeled as a flexible string. Schematic representations.*

*Fig. 2: Collisions with neighboring lipids are modeled by self-consistent confining potential. Potential is parabolic in the string deviation from axis z (arrows size mimics local force strength).*

Boundary conditions for a model flexible string take into account the following physical assumptions (same is assumed also for component $R_y(z)$):

$$\begin{aligned}R'_x(0) &= 0 &&\text{a chain terminates perpendicularly to the membrane surface}\\ R'''_x(0) &= 0 &&\text{net force acting on a head is zero}\\ R''_x(L) &= 0 &&\text{net torque acting at a chain's free end is zero}\\ R'''_x(L) &= 0 &&\text{net force acting at a chain free end is zero}\end{aligned} \quad (A.2)$$

The first boundary condition reflects the orientational asymmetry of the monolayer due to water-lipid interface which is clearly seen from data on the molecules orientational order parameter [4,5]: lipid tails are more ordered in the vicinity of headgroups constrained by the hydrophobic tension. Yet, the chains are not permanently perpendicular to the membrane surface, and boundary conditions are approximate and necessary to keep the model analytically solvable.

Assuming membrane to be locally isotropic in the lateral plane, partition function can be split in the product of the two equal components, $Z = Z_x Z_y = Z_x^2$, and thus free energy of the lateral oscillations of the chain equals to:

$$F_t = -2k_B T \log Z_x \quad (A.3)$$

Partition function $Z_x$ could be written as a path integral over all chain conformations:

$$Z_x = \iint e^{-\frac{E\{R_x(z), \dot{R}_x(z)\}}{k_B T}} DR_x D\dot{R}_x \quad (A.4)$$

Under the boundary conditions (A.2) potential part of the energy functional (A.1) can be equivalently rewritten in terms of linear Hermitian operator $\hat{H} = B + K_f \frac{\partial}{\partial z^4}$ in the form:

$$E_{t(pot)} = \sum_{\alpha=x,y} E_\alpha$$

$$E_\alpha \equiv \int_0^L \left[ R_\alpha(z) \hat{H} R_\alpha(z) \right] \, dz \tag{A.5}$$

Then, an arbitrary conformation of the chain is expressed as the deviation of the centers of the string, $R_{x,y}(z)$, from the straight vertical line (see Fig 1), and is parameterized by an infinite set of coefficients $C_n$ of the linear decomposition of the function $R_{x,y}(z)$ over the eigenfunctions $R_n(z)$ of the operator $\hat{H}$:

$$R_{\alpha=x,y}(z) = \sum_n C_{n,\alpha} R_n(z)$$

$$\hat{H} R_n(z) = E_n R_n(z) \tag{A.6}$$

Substituting Eq. (A.6) into Eq. (A.1) and using the standard orthogonality property of the eigenfunctions of operator $\hat{H}$, enables a simple decomposition of the energy functional into the series:

$$E_t = \sum_n \frac{1}{2} \left\{ \rho \dot{C}_n^2 + E_n C_n^2 \right\} \tag{A.7}$$

We thus see that energy of a fluctuating string in a parabolic potential maps on the sum of energies of harmonic oscillators with rigidities $E_n$. Hence, the Boltzmann's probability of the state of a string in an arbitrary conformation $R_{x,y}(z)$, $P(\{R_{x,y}(z)\})$, is proportional to the infinite product of the Boltzmann's probabilities of the states of these oscillators due to obvious relation:

$$P(\{R_{x,y}(z)\}) \propto \exp\left\{-\frac{E_t}{k_B T}\right\} \sim \prod_n \exp\left\{-\frac{\epsilon_n}{k_B T}\right\}$$

$$\epsilon_n \equiv \frac{1}{2} \left\{ \rho \dot{C}_n^2 + E_n C_n^2 \right\} \tag{A.8}$$

Therefore, distribution of the coefficients $C_n$ prove to be just Gaussian Boltzmann's distribution, which makes the whole thermodynamic theory of the lipid membrane analytically tractable. The corresponding eigenvalues $E_n$ and eigenfunctions $R_n(z)$ of the operator $\hat{H} = B + K_f \frac{\partial}{\partial z^4}$ are [6]:

$$n = 0 \quad \Rightarrow \quad \begin{cases} E_0 = B \\ R_0(z) = \sqrt{1/L} \end{cases}$$

$$n \in \mathbb{N} \quad \Rightarrow \quad \begin{cases} c_n = \pi n - \frac{\pi}{4} \\ E_n = B + c_n^4 \frac{K_f}{L^4} \\ R_n(z) = \sqrt{\frac{2}{L}} \left[ \cos(c_n \frac{z}{L}) + \frac{\cos(c_n)}{\cosh(c_n)} \cosh(c_n \frac{z}{L}) \right] \end{cases} \tag{A.9}$$

This gives the following product of the Gaussian integrals for the partition function:

$$Z_x = \int\limits_{-\infty}^{+\infty} \prod_n e^{-\frac{(\rho \dot{C}_n)^2}{2\rho k_b T} - \frac{C_n^2 E_n}{2 k_B T}} \frac{\mathrm{d}(\rho \dot{C}_n) \cdot \mathrm{d}C_n}{2\pi \hbar} = \prod_n \frac{k_B T}{\hbar} \sqrt{\frac{\rho}{E_n}} \qquad (A.10)$$

## B. Derivation of chain's constant contour length condition

Assuming chain's full (contour) length, $L_R$, is constant one has:

$$L_R = \int_0^L \sqrt{1 + \left\langle \left(\frac{\partial \vec{R}}{\partial z}\right)^2 \right\rangle} \, \mathrm{d}z \qquad (B.1)$$

Our goal here is to express chain's length $L$ in terms of full length $L_R$ and other properties of the string. To that end we evaluate the integral in Eq. (B.1).

One can re-write the mean value under the square root:

$$\vec{R} = R_x \vec{e}_x + R_y \vec{e}_y$$
$$\frac{\partial \vec{R}}{\partial z} = \frac{\partial R_x}{\partial z} \vec{e}_x + \frac{\partial R_y}{\partial z} \vec{e}_y$$
$$\left(\frac{\partial \vec{R}}{\partial z}\right)^2 = \left(\frac{\partial R_x}{\partial z}\right)^2 + \left(\frac{\partial R_y}{\partial z}\right)^2$$
$$\left\langle \left(\frac{\partial \vec{R}}{\partial z}\right)^2 \right\rangle = \left\langle \left(\frac{\partial R_x}{\partial z}\right)^2 \right\rangle + \left\langle \left(\frac{\partial R_y}{\partial z}\right)^2 \right\rangle \equiv 2 \left\langle \left(\frac{\partial R_x}{\partial z}\right)^2 \right\rangle \qquad (B.2)$$

Substituting Eq. (B.2) into Eq. (B.1) leads to

$$L_R = \int_0^L \sqrt{1 + 2\left\langle \left(\frac{\partial R_x}{\partial z}\right)^2 \right\rangle} \, \mathrm{d}z$$
$$\approx \int_0^L \left(1 + \left\langle \left(\frac{\partial R_x}{\partial z}\right)^2 \right\rangle\right) \mathrm{d}z$$
$$= L + \int_0^L \left\langle \left(\frac{\partial R_x}{\partial z}\right)^2 \right\rangle \mathrm{d}z \qquad (B.3)$$

The mean value under the integral is evaluated using the following relations:

$$R_x = \sum_n C_n R_n$$
$$R'_x = \sum_n C_n R'_n$$
$$(R'_x)^2 = \sum_{n,m} C_n C_m R'_n R'_m$$
$$\langle (R'_x)^2 \rangle = \sum_{n,m} \langle C_n C_m \rangle R'_n R'_m = \sum_n \frac{k_B T}{E_n} (R'_n)^2 \qquad (B.4)$$

Upon calculating $R'_n$ (see Eq. (9) in the manuscript), substituting Eq. (B.4) into (B.3) and introducing dimensionless coordinate $x = z/L$, one arrives at quadratic equation for $L$:

$$L_R = L \left( 1 + \frac{2 k_B T L}{K_f} \sum_{n=1} \frac{c_n^2 \int_0^1 (r'_n)^2 \, dx}{b + c_n^4} \right) \quad (B.5)$$

Solving the above equation yields Eq. (12) of the manuscript.

## C. Renormalization of chain's free energy

It is clear that free energy of flexible string diverges (see Eqs. (3, 10, 9) in the manuscript). This has never been a problem in our previous works, where we was only concerned with derivatives of free energy [6–8], but in this work we need to plot free energy of the string. For this we re-normalize free energy of the flexible string by the free energy of the rigid string [8].

To that end, consider a non-trivial Jacobian, $C$, involved in $Z_x$ calculation:

$$Z_x = \frac{1}{C} \prod_n \frac{k_B T}{\hbar \omega_n} \quad \Rightarrow \quad F_t = -2 k_B T \sum_n \log \left( \frac{k_B T}{\hbar \omega_n} \right) + 2 k_B T \log C \quad (C.1)$$

(see Eqs. (9,10) in the manuscript). Free energy Eq. (C.1) can be re-written in the following form:

$$F_t = 2 k_B T \log \left( \frac{\hbar \sqrt{B/\rho}}{k_B T} \right) + 2 k_B T \log C +$$

$$+ k_B T \sum_{n=1} \log \left( 1 + \frac{B L^4}{c_n^4 K_f} \right) + 2 k_B T \sum_{n=1} \log \left( \frac{\hbar \sqrt{c_n^4 \frac{K_f}{\rho L^4}}}{k_B T} \right) \quad (C.2)$$

Considering rigid rod, $K_f \to \infty$, we can obtain its free energy as a limiting case of Eq. (C.2):

$$F_t^{\text{rigid}} = 2 k_B T \log \left( \frac{\hbar \sqrt{B/\rho}}{k_B T} \right) + 2 k_B T \log C + 2 k_B T \sum_{n=1} \log \left( \frac{\hbar \sqrt{c_n^4 \frac{K_f}{\rho L^4}}}{k_B T} \right) \quad (C.3)$$

But from the other hand, rigid rod has a single oscillation mode, $\omega_0$, and we know its partition function exactly:

$$Z_x^{\text{rigid}} = \frac{k_B T}{\hbar \omega_0} \quad \Rightarrow \quad F_t^{\text{rigid}} = -2k_B T \log \left( \frac{\hbar \sqrt{B/\rho}}{k_B T} \right) \tag{C.4}$$

Comparing Eq. (C.4) with Eq. (C.3) one concludes that

$$\log C = - \sum_{n=1} \log \left( \frac{\hbar \sqrt{c_n^4 \frac{K_f}{\rho L^4}}}{k_B T} \right) \tag{C.5}$$

and hence free energy of the flexible strings steric repulsion renormalized by the free energy of the rigid string (rod) is :

$$\frac{\tilde{F}_t}{k_B T} = 2 \log \frac{\hbar \sqrt{\frac{B}{\rho}}}{k_B T} + \sum_{n=1} \log \left( 1 + \frac{B L^4}{K_f c_n^4} \right) \tag{C.6}$$

## D. Eq. (12) in the manuscript

Introducing dimensionless $b$ via:

$$B = b \frac{K_f}{L^4}, \tag{D.1}$$

one can re-write Eq. (C.6) in the form:

$$\frac{\tilde{F}_t}{k_B T} = 2 \log \frac{\hbar \sqrt{\frac{K_f}{\rho L^4}}}{k_B T} + \log(b) + \sum_{n=1} \log \left( 1 + \frac{b}{c_n^4} \right), \tag{D.2}$$

The sum, $\log(b) + \sum_{n=1} \log \left( 1 + \frac{b}{c_n^4} \right)$, can be evaluated using technique of differentiation by parameter:

$$\Sigma = \log b + \sum_{1}^{\infty} \log \left( 1 + \frac{b}{c_n^4} \right) \quad \Rightarrow \quad \frac{\partial \Sigma}{\partial b} = \frac{1}{b} + \sum_{n=1}^{\infty} \frac{1}{b + c_n^4} \approx \frac{3}{4b} + \frac{1}{2\sqrt{2} b^{3/4}}$$
$$\Rightarrow \Sigma = \int \left( \frac{3}{4b} + \frac{1}{2\sqrt{2} b^{3/4}} \right) db = \frac{3}{4} \log b + \sqrt{2} b^{1/4} + \text{const} \tag{D.3}$$

Here const does not depend on $b$. Substituting $B$ back using Eq. (D.1) one arrives at Eq. (12) of the manuscript:

$$\frac{\tilde{F}_t}{k_B T} = 2 \log \frac{\hbar \sqrt{\frac{K_f}{L^4 \rho}}}{k_B T} + \frac{3}{4} \log \left( L^4 \frac{B}{K_f} \right) + \sqrt{2} L \sqrt[4]{\frac{B}{K_f}}. \tag{D.4}$$

## E. Stability of phase transition temperature

We analyze stability of the model predictions for transition temperature $T_m$ against small changes of the values of the model parameters used in the paper: $K_f$, $A_n$, and VdW coupling constant $U$. The results are plotted below:

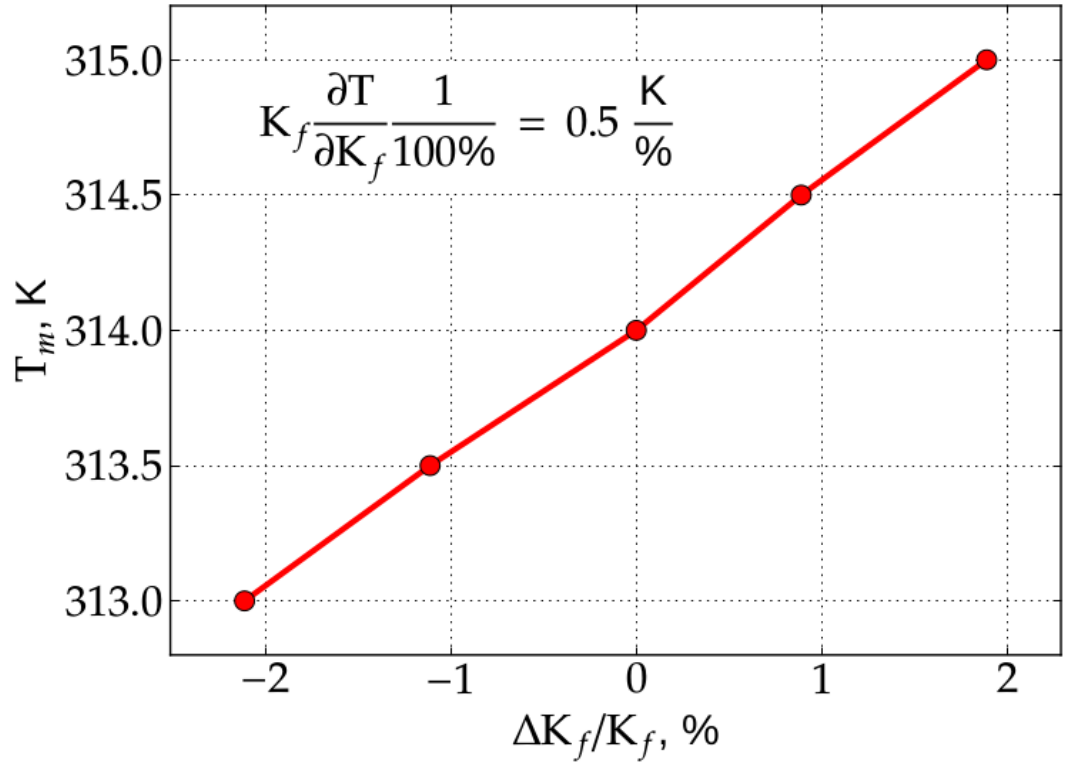

*Fig. 3: Dependence of the temperature of the main phase transition on the bending rigidity of the string relative to the reference value used in the paper. Changing bending rigidity by one percent results in change of $T_m$ by 0.5 K, i.e 0.15 %.*

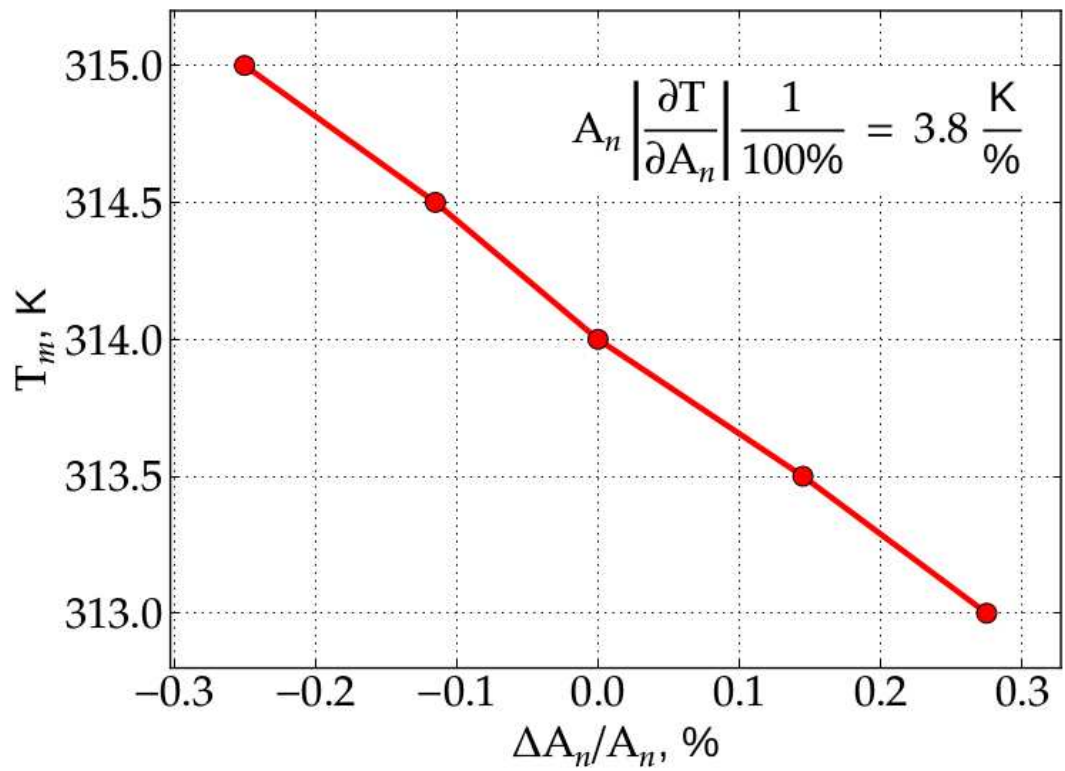

*Fig. 4: Dependence of the temperature of the main phase transition on the incompressible area of the lipid relative to the reference value used in the paper. Changing incompressible area by one percent results in change of $T_m$ by 4 K, i.e 1.2 %.*

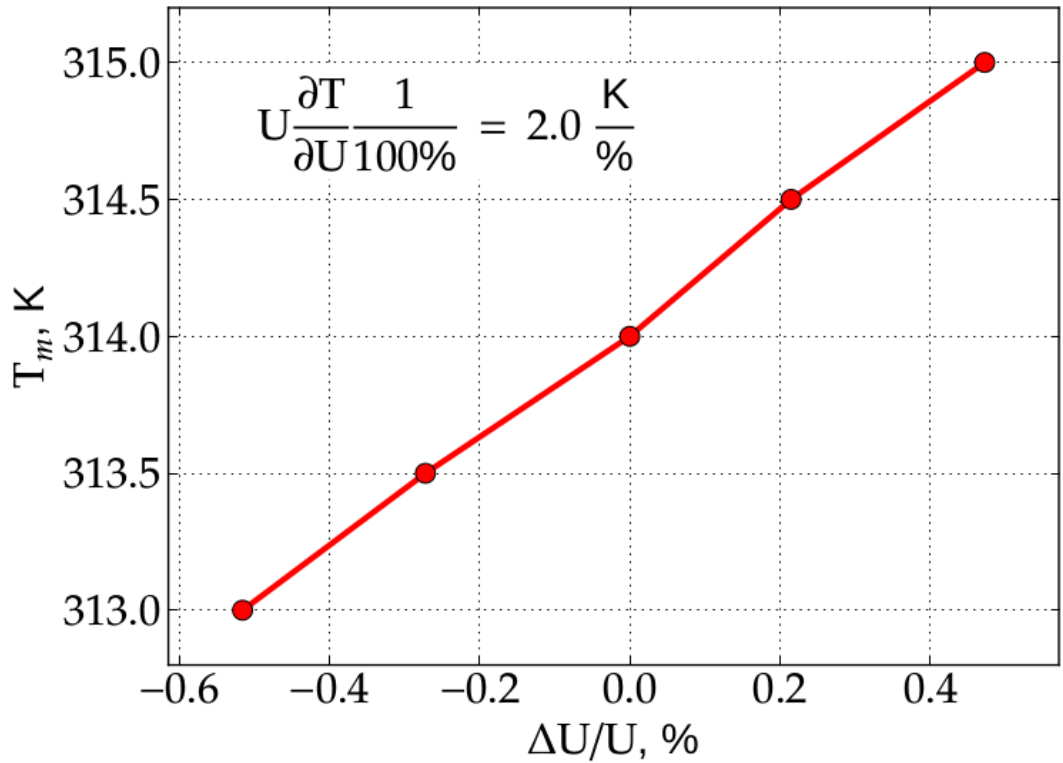

*Fig. 5: Dependence of the temperature of the main phase transition on the van der Waals interaction parameter relative to the reference value used in the paper. Changing van der Waals parameter by one percent results in change of $T_m$ by 2 K, i.e 0.6 %.*

So, we have demonstrated that model is quite stable against small changes of the model parameters: separate changes of these parameters by one percent relative to the reference values results in respective changes of $T_m$ by relative amounts of 0.15%, 1.2% and 0.6% respectively. As follows from these results, our model turns out to be most sensitive to the change of the incompressible area, $A_n$. This happens ultimately due to a high power to which the area per lipid enters the van der Waals interaction expression.